\documentclass[]{raa}            
\usepackage{graphicx,times}
\usepackage{natbib}

\providecommand{\bm}[1]{\mbox{\boldmath$#1$\unboldmath}}
\def\kms{$\rm{km~s}^{-1}$}

\begin{document}

   \title{Kinematics and Stellar Population Properties of the Andromeda Galaxy by the Spectroscopic Observations of the Guoshoujing Telescope
}
\volnopage{ {\bf 20xx} Vol.\ {\bf 9} No. {\bf XX}, 000--000}
   \setcounter{page}{1}

   \author{Hu Zou
      \inst{1,2}
   \and Yanbin Yang
      \inst{1}
   \and Tianmeng Zhang
      \inst{1}
   \and Jun Ma
      \inst{1}
   \and Xu Zhou
      \inst{1}
   \and Ali Luo
      \inst{1}
   \and Haotong Zhang
      \inst{1}
   \and Zhongrui Bai
      \inst{1}
   \and Yongheng Zhao
      \inst{1}
   }

   \institute{National Astronomical Observatories, Chinese Academy of Sciences,
             Beijing 100012, China; {\it zhouxu@bao.ac.cn}\\
        \and
             Graduate University of Chinese Academy of Sciences, Beijing 100049, China
\vs \no
   {\small Received [year] [month] [day]; accepted [year] [month] [day] }
}

\abstract{The Andromeda galaxy was observed by the Guoshoujing Telescope (GSJT, formly named the Large Sky Area Multi-Object Fiber
Spectroscopic Telescope -- LAMOST) during the 2009 commissioning phase. Due to the absence of standard stars for flux calibration,
we use the photometric data of 15 intermediate bands in the Beijing-Arizona-Taipei-
Connecticut (BATC) survey to calibrate the spectra. Total 59 spectra located in the bulge and disk of the galaxy are obtained.
Kinematic and stellar population properties of the stellar content are derived with these spectra. We obtain the global velocity
field and calculate corresponding rotation velocities outer to about 7 kpc along the major axis. These rotation velocity measurements
complement those of the gas content, such as the H {\sc i} and CO. The radial velocity dispersion presents that the stars in the bulge
are more dynamically thermal and the disk is more rotation-supported. The age distribution shows that the bulge was formed about 12 Gyr
ago, the disk is relatively younger, and the ages of some regions along the spiral arms can reach as young as about 1 Gyr. These
young stellar populations have relatively richer abundance and larger reddening. The overall average metallicity of the galaxy
approximates the solar metallicity and a very weak abundance gradient is gained. The reddening map gives a picture of a dust-free
bulge and a distinct dusty ring in the disk.
}

\authorrunning{Zou et al.}            
\titlerunning{Kinematics and Stellar Population Properties of M31}  
\maketitle


%
%
\section{Introduction}           
Nearby galaxies are good probes to investigate the kinematic and dynamical properties, stellar
populations, and formation and evolution histories of present-day galaxies in the local universe. Detailed
analyses can be performed in respect to the bulge, disk, spiral arms, dust, H {\sc ii}
regions and so on with imaging of high resolutions and spectroscopy of high quality other than
individual stars in the Milk Way. Therefore, large samples of nearby galaxies, covering different
morphological types, luminosities, star formation rates and other properties, were handpicked
in a variety of photometric and spectroscopic surveys from radio (\citealt{is83, he03, wa08}), ultraviolet
(\citealt{gi07}), optical(\citealt{ke92, ro10}) to infrared (\citealt{ke03}) wavebands in order to study
the chemical abundance, stellar population, stellar formation rate, gas content, interstellar medium
and other physical properties of galaxies.

Most of the spectroscopic observations about the nearby galaxies were focused on the bright galactic
cores (\citealt{he80, bo95}) and large H {\sc ii} regions (\citealt{ne76, va98}), whose spectra
were obtained mainly by aperture or long-slit spectrographs. A few large-aperture spectroscopic
observations of nearby galaxies, such as \cite{ga89} and \cite{ke92}, were taken in order to gain
the integrated spectra and compare their integrated features with those of distant galaxies.
Recently, integrated field spectroscopy techniques have played a very important role on spatially
resolved spectroscopic measurements of nearby galaxies with high spatial resolutions in substantially large
field of views (\citealt{ba01, ro10}).

The Andromeda galaxy (M31 or NGC 224), as the largest and nearest spiral galaxy (SA(s)b) in the
Local Group, has a major diameter of about 190{\arcmin} and distance of about 784 kpc (\citealt{st98}).
The systemic velocity is about -300 km s$^{-1}$, the inclination is about 78$^\circ$ and the position
angle of the major axis is about 38$^\circ$ (\citealt{go70}). These characteristic parameters will be
adopted throughout the following study of this paper. Due to its great apparent scale length, distinct
components (e.g., bulge, disk, spiral arms and halo) and a considerable number of resolvable sources
and substructures such as single stars, star clusters, H {\sc ii} regions, dust lanes and gas ingredients
can be explored by imaging and spectroscopy to understand its distance, stellar population, age, chemical
abundance distributions, tidal interaction, galactic formation and dynamical evolution
(\citealt{va69, bl82, br84, ch02, ib05, ch08}).

Due to its large optical size (optical radius $R_{25} = 1.59^\circ$), M31 is chosen to be one of the
testing targets during the commissioning of the GSJT survey
whose field of view (FOV) is about 5 degrees
(\citealt{wa96, su98, cu09}). The GSJT, located at the Xinglong Station of National Astronomical
Observatories of China, is a quasi-meridian reflecting Schmidt telescope with a clear aperture
of 4 meters and a focal length of 20 m. Special and unique designs make this telescope possess
both large aperture and FOV. Four thousand fibers deployed on the focal plane of a 1.75 m diameter
can simultaneously obtain the spectra of 4000 celestial objects. Individual point sources, normal
and luminous red galaxies, planetary nebulae (PNe) and candidates as well as low red-shift quasars
are selected as the observed objects in two M31 testing fields which are centered close to the
optical nucleus of M31 and in the northeastern halo, respectively.

Although the GSJT is undergoing the commissioning phase and many aspects such as the telescope observing
conditions (e.g., dome seeing), fiber positioning, instruments and data processing software are not in
their perfect status, some preliminary results were published with the commissioning data observed in
the winter of 2009 (quasars of \cite{hu10}, \cite{wu10a}, and \cite{wu10b} and planetary nebulae of \cite{yu10}).
At present, the pointing of fibers can not be directed so accurately, but they lie not far away from their
preassigned positions on the sky. For studying the galactic surface, the accuracy of pointing is not so
important because the fibers still point to the places of our interest. Observing conditions like seeing
are not important at all for the extended galaxy. We will make use of the spectra near the core and inner disk
in one of the M31 testing fields to study the distributions of the radial velocity, velocity dispersion,
age, metallicity and reddening. In view of the absence of appropriate standard stars, the photometric
observations of 15 intermediate-band filters in the BATC survey (\citealt{fa96}) are used to flux-calibrate
the GSJT spectra.

There are substantial published studies about the kinematics and stellar population properties of M31, which
can be used to compare with our results. Both the neutral atomic hydrogen (H {\sc i}) observation by the
Half-Mile telescope with an angular resolution of 1.5--2.2{\arcmin} and a velocity resolution of 39 km s$^{-1}$
(\citealt{em76}) and a molecule $^{12}$CO(J=1--0)-line survey with a high angular resolution of 23{\arcsec}
and a velocity resolution of 2.6 km s$^{-1}$ (\citealt{ni06}) mapped the radial velocity field of the Andromeda
galaxy. A large velocity gradient, rotational characteristics and much lower gas density near the center than
the outer gas ring were presented in these two measurements. A slit spectrum covering the wavelength range from
4918 to 5302 {\AA} at the point-like nucleus of M31 was obtained by \cite{mo73}. Comparing the spectrum with
those of G and K stars, they gained the total line of sight velocity dispersion of about 120 $\pm$ 30 km s$^{-1}$.
\cite{wh80} also observed the spectrum of the nucleus and provided the nuclear velocity dispersion of about
181 km s$^{-1}$. Meanwhile, dispersion measurements in some literature were summarized by him, giving
the average velocity dispersion of about 164 km s$^{-1}$.

Multi-band photometric data from ultraviolet to infrared and images with resolved stars from the Hubble
Space Telescope were used to drive the age, mass, metallicity and reddening of globular clusters in M31 by either
stellar population synthesis or color magnitude diagrams (\citealt{fa08,ma09,pe10}). Spectroscopic data of several
hundred of globular clusters extending from the galactic center out to 1.5$^\circ$ revealed that metal-richer
clusters have a centrally concentrated distribution with high rotation amplitudes, which are consistent with
the bulge population (\citealt{pe02}). These globular clusters present panoramic views of different parameters
and provide extremely good constraints of the structure formation and evolution of this spiral galaxy. Large
samples of planetary Nebulae, supernova remnants and H {\sc ii} regions were also observed to obtain the chemical
abundance and reddening features across the whole galaxy (\citealt{ku79,bl82,ja99,ga99,ri99}).

This paper is organized as follows. In \S \ref{sec-obs}, the telescopes, relevant facilities and observations
are described. In \S \ref{sec-data}, we present the detailed data reduction including the data processing
of the GSJT spectra and flux-calibrating these spectra using the photometric observations of 15 intermediate bands
in the BATC survey. Results of the kinematic and stellar population parameters derived by the synthesis model
and discussions on the distributions of these parameters are given in \S \ref{sec-dis}. Finally, conclusions
are summarized in \S \ref{sec-conc}.

\section{Telescopes and Observations} \label{sec-obs}
The GSJT is the largest optical reflecting Schmidt telescope in China with a clear aperture of 4 meters
and a wide FOV of 5 degrees (\citealt{wa96, su03}). Its architecture is comprised of a reflecting Schmidt
corrector (M$\rm_{A}$) at the north end, a spherical primary mirror (M$\rm_{B}$) at the south end and
a focal plane with diameter of 1.75 meters lying between M$\rm_{A}$ and M$\rm_{B}$ (20 meters away
from the primary mirror). Light from celestial objects is reflected by M$\rm_{A}$ into the telescope
enclosure, reflected again by M$\rm_{B}$ and at last converged at the focal plane. M$\rm_{A}$, as a coelostat
and corrector, consists of 24 hexagonal plane submirrors giving the size of 5.72$\times$4.40 m$^2$. It is
exposed to air with a wind screen preventing wind buffeting. This alt-azimuth mounting corrector together
with the focal plane can track the celestial objects within the sky area of $-10^\circ < \delta < 90^\circ$
for 1.5 hours during their crossing the meridian. M$\rm_{B}$ has 37 spherical submirrors giving the size
of 6.67$\times$6.05 m$^2$ and its enclosure is ventilated to keep a good dome seeing. Active optics composed
of segmented mirror active optics and thin deformable mirror active optics is achieved by the force actuators
set on each submirror and two wave front sensors (\citealt{su04}).

At the focal plane of the telescope, there are 4000 fibers medially distributed to 16 spectrographs. Each
fiber with a diameter of 3.3 arcsec is directed to an entrance slit, where a dichroic filter splits the
incident light beam into blue (3700--5900\AA) and red (5700--9000\AA) channels. Low and median resolution
spectrographs are or will be equipped and in the current phase only the low resolution spectrographs come
into service. Full and half sizes of slits yield the spectral resolutions of 1000 and 2000. All the fibers
are parallelly controlled by two micro-stepping motors in individual domains (within a circle of 30 mm diameters,
corresponding to 340 arcsec) (\citealt{xi98}). The separation between two adjacent units is designed to be
20 mm which ensures no blind spot on the convex focal surface. The minimal separation of two observable
objects is 55{\arcsec} in order to avoid the collision of fibers.

During the early commissioning phase, nine fields were proposed to be observed in the winter of 2009, which
is the best observational season in the Xinglong Station (\citealt{yu10}). Two fields related to M31 were
included in the test plan: one is centered on $\alpha = 11^\circ.155, \delta = 40^\circ.679$, close to
the optical galactic center ($\alpha = 10^\circ.685, \delta = 41^\circ.269$), and the other is located in
the northeastern outskirt of the Andromeda halo, centered on $\alpha = 18^\circ.142, \delta = 45^\circ.338$.
Sorts of objects were selected as targets including low redshift quasars, normal and luminous red galaxies,
standard stars, PNe and PN candidates, and sources complemented by the stars from the Two Micron All Sky
Survey. Nearly 5\% of the fibers were assigned to get the sky spectra. Observations were carried out on 2009
October 19 and December 15 (see details in \cite{hu10}). Bright and faint sources are observed respectively
with three exposures (600 s each) and two exposures (1800 s each). Arc spectra from a mercury vapor and neon lamp and
flat fields from a tungsten lamp and the offset sky were also obtained during observation. Given the goals
of this paper and the spectral quality, only the field centered near the nucleus is analyzed in the following
sections.

Since there is no available spectroscopic standard star in the field of our interest, we intend to flux-calibrate
all the spectra by the photometric data of 15 intermediate bands in the BATC survey. The survey is based on a 60/90
cm Schmidt telescope also mounted at the Xinglong Station (\citealt{fa96}). A 2048$\times$2048 charge-coupled
device (CCD) with the pixel size of 1.7{\arcsec} is installed at the focal plane (focal ratio: f/3), generating
a FOV of 58{\arcmin}$\times$58{\arcmin}. A set of 15 intermediate-band filters with bandwidths of 200--300 {\AA}
covers the wavelength range from 3000 to 10000 {\AA}. They are specifically designed to exclude the bright and
variable night sky emission lines. Four Oke-Gunn standard stars are used for flux calibration during the photometric
nights (\citealt{ya00,zh01}). In the BATC survey project, dozens of nearby galaxies including the Andromeda galaxy
have been observed. Observations centered on the galactic nucleus of M31 began in 1995 and all the normal and
photometric observations for 15 filters are completed now.

\section{Data Reduction} \label{sec-data}
\subsection{Normal Data Processing for the GSJT}
The original blue and red CCD frames are separately reduced by the basic two-dimensional data processing
pipeline \citep{lu04}. In the beginning, CCD biases are subtracted from the raw frames including arc lamp,
flat and target ones. Each spectral frame includes 250 spectra and each spectrum covers about 16 pixels
in the spatial direction. Because of the high signal to noise (S/N), flat flames are used to trace the fiber
spectra. The profile centers are determined by centroids and a proper aperture is selected to extract the
flux along the dispersion direction for each fiber. Then, the extracted one-dimensional target spectra are
corrected for various variations (e.g., relative throughputs of fibers, fiber transmissions and spectrograph
responses) by the the lamp and sky flat fields. Wavelength calibrations are performed with the help of the
arc lamp and sky emission lines which enforce the calibration accuracy.  All sky spectra are combined to
construct a ``super sky'', which is interpolated to remove the sky light from the specified target spectra.
Due to the absence of appropriate standard stars in the M31 field, we can get reliable blue and red spectra
only with wavelength calibrations and sky subtractions but without flux calibration.

\begin{figure}
\centering
\includegraphics[width=\textwidth, angle=0]{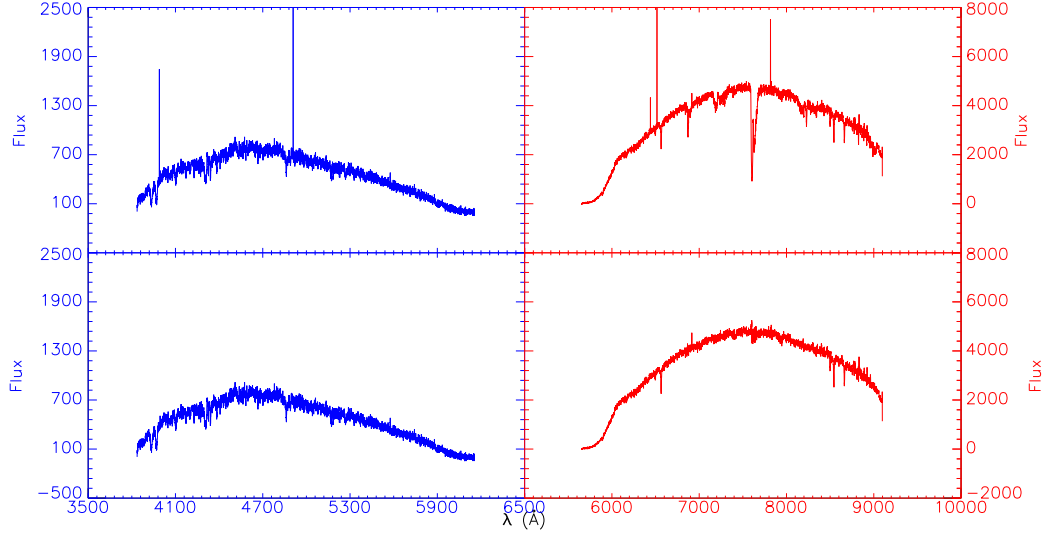}
\begin{minipage}[]{\textwidth}
\caption{Top: Blue and red spectra of a bright star in its original form. Bottom: The same blue spectra after removing the cosmic rays
and the same red spectrum after removing the cosmic rays and correcting the atmospheric absorption lines.\label{fig1}}
\end{minipage}
\end{figure}

Cosmic rays are removed via replacing the relevant pixel values by interpolations. The red spectrum has considerable
strong atmospheric absorption lines from O$_2$ and H$_2$O around 6300 {\AA}, 6900 {\AA}, 7200 {\AA}, 7600 {\AA},
8200 {\AA} and 9400 {\AA}, but the blue spectrum is to the contrary. We select several spectra of relatively bright stars who
have few intrinsic absorption lines in their red wavelength region and use them to construct a normalized spectrum of
the atmospheric absorption. All observed red spectra are corrected by this normalized spectrum. Figure \ref{fig1}
displays the results of removing cosmic rays for the blue spectrum of a bright star in the left part and removing cosmic
rays and correcting atmospheric absorption lines for the red spectrum of the same star. Finally, spectra of different
exposures are combined to reduce noises.

\subsection{Flux Calibration by the BATC Photometric Data}
Photometric CCD images from the BATC survey are reduced by some standard procedures including the bias subtraction,
flat-fielding and astrometry. Frames of different exposures for each band are combined to increase the S/N and delete the
cosmic rays. Meanwhile, frames with relatively large sky backgrounds or short exposures are eliminated. Standard stars
observed at photometric nights are used to obtain the instrumental zeropoints and atmospheric extinctions, which
can convert the CCD counts to the out-of-atmosphere fluxes.

\begin{figure}
\centering
\includegraphics[width=\textwidth, angle=0]{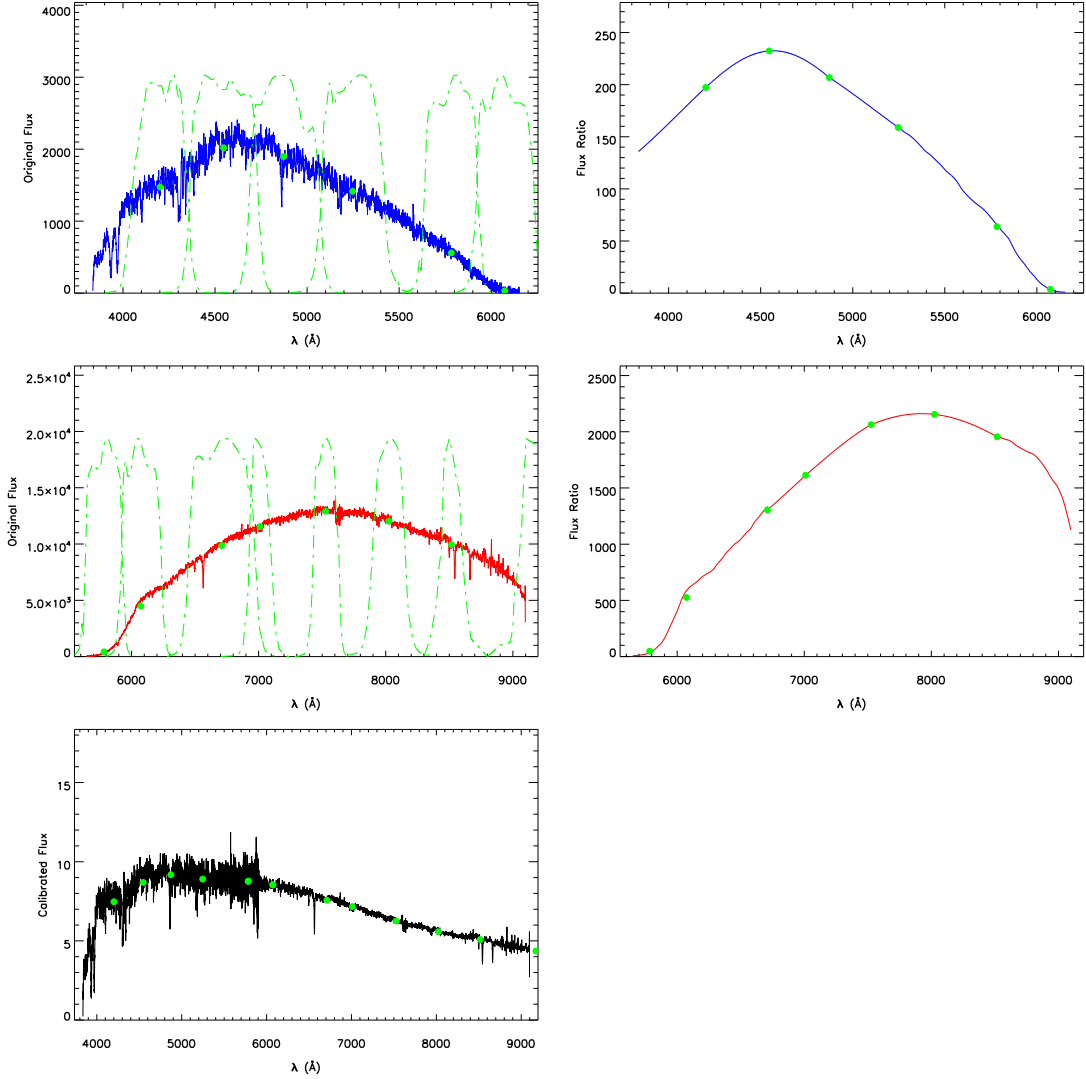}
\begin{minipage}[]{\textwidth}
\caption{Top left: Blue spectrum of the same star as shown in Figure \ref{fig1} and convolutions with the BATC
filter transmissions as plotted in green dash-dotted curves. The green filled circles are the convolved
instrumental fluxes. Top right: Flux ratios between the GSJT instrumental fluxes and the BATC calibrated fluxes. Green
filled circles are the ratios of 6 bands and the curve in blue is the interpolated instrumental response. Middle
panels present the same contents for the red spectrum except for convolutions with 8 bands. Bottom: Flux-calibrated
and combined spectra in solid curves derived by the BATC photometric data in green circles. Flux in the left panel is
in $10^{-17}$ ergs s$^{-1}$ cm$^{-2}$ {\AA}$^{-1}$ fiber$^{-1}$.\label{fig2}}
\end{minipage}
\end{figure}

Since the M31 sky field has been observed by the BATC survey using 15 intermediate band filters, for the GSJT fibers
near the galactic core and bright disk we can flux-calibrate the spectra of these fibers very well with the spectral
energy distributions (SEDs) in 15 discrete wavelengths, ranging from 3000 {\AA} to 10000 {\AA}. First, we map all
the objects in the GSJT input catalog to the BATC images and calculate the calibrated fluxes in the fiber aperture
of 3.3{\arcsec}. Actually, the astrometry of photometric images has errors, seeing condition affects the flux
distributions and the fiber-positioning is not so accurate. So we count the fluxes in a diameter of 6 pixels in the
BATC images and then scale them to the fiber size of 3.3{\arcsec}. Second, we convolve the LAMOST spectra with the
BATC filter transmission curves to obtain the instrumental fluxes. Both the blue and red spectra and convolutions
with the filter transmissions for the same star as shown in Figure \ref{fig1} are displayed in the top-left and
middle-left panels of Figure \ref{fig2}. The blue spectrum is convolved with 6 filter transmissions (green dash dotted
curves) ranging from 4200 to 6075{\AA} and the red one is convolved with 8 filter transmissions ranging from 5795 to
9190{\AA}. Flux ratios between the convolved instrumental fluxes and calibrated fluxes, as shown in the filled circles
in the top-right and mid-right panels of Figure \ref{fig2}, are computed to derive the instrumental responses.
Due to the complexity of instrumental responses near both two ends of the spectra, we interpolate the calibrated flux
at several interm wavelengths among the bands in these areas, then get the corresponding fluxes in the original
spectra at the same positions and calculate their ratios. The continuous instrumental responses are obtained by
the piecewise quadratic interpolation using all the discrete flux ratios as plotted as curves in those two panels.
These response curves are applied to flux calibrate the original spectra. At last, the red and blue calibrated spectra
are spliced into a single spectrum as shown in the bottom panel of Figure \ref{fig2}.

\begin{figure}
\centering
\includegraphics[width=\textwidth, angle=0]{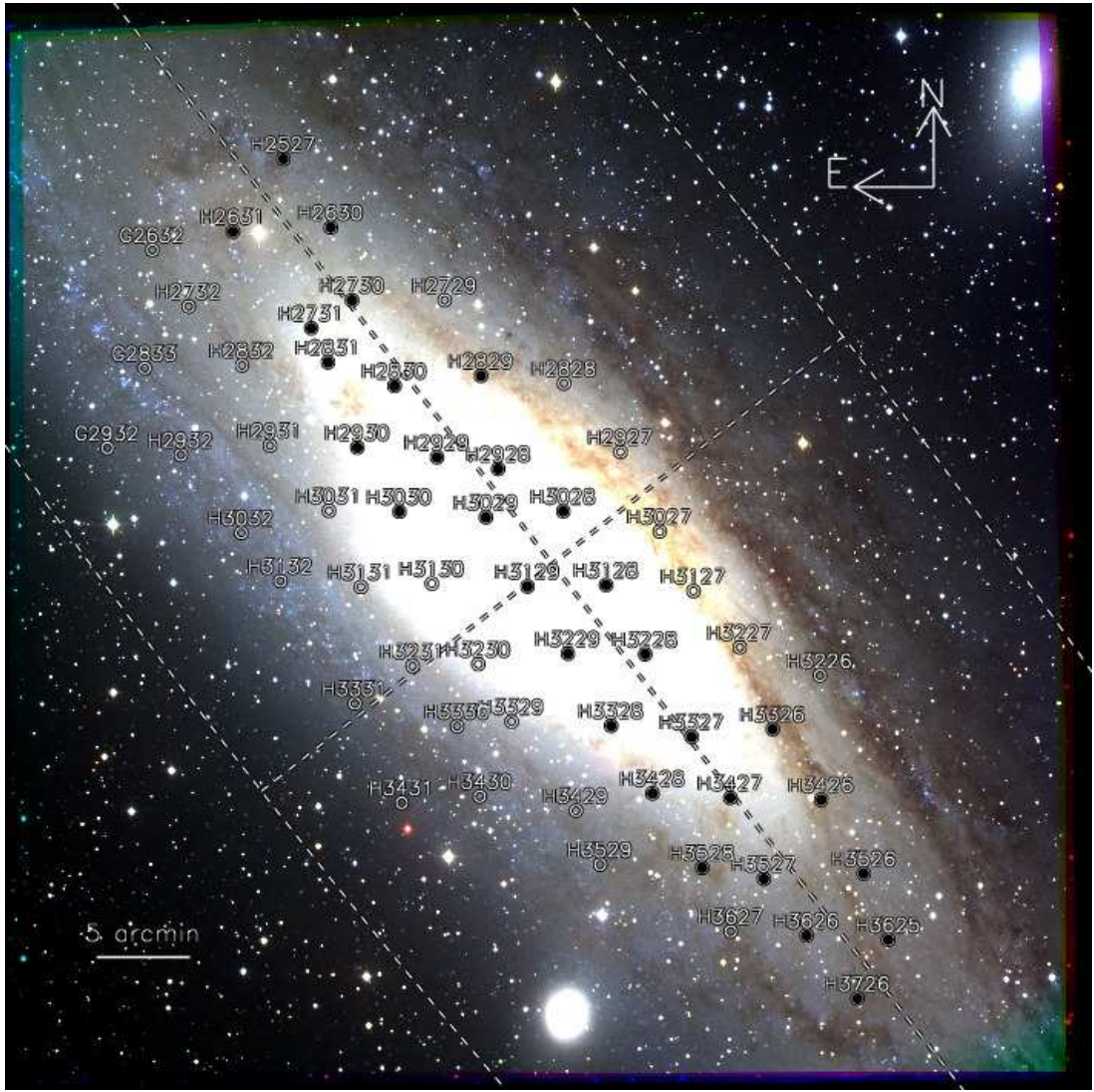}
\begin{minipage}[]{\textwidth}
\caption{True color map of M31 combined with three intermediate band images whose effective wavelengths are 4550, 6075 and 8082 {\AA}.
Circles marked by their temporary IDs as presented in Table \ref{tab1} are the usable spectra observed by the GSJT. The dashed
arcs are the clipped elliptical enclosure of the Andromeda's disk. The radius of the major axis is about 1.59$^\circ$, the position
angle of the major axis is 38$^\circ$ and the disc inclination angle is 78$^\circ$. Two perpendicular dashed lines are the optical
major and minor axis. Filled circles are the positions, whose projected distances away from the major axis are within 1 kpc
($\sim$4.4 {\arcmin}). These positions are chosen to investigate of the radial distributions of calculated properties as will be shown in \S \ref{sec-dis}.
\label{fig3}}
\end{minipage}
\end{figure}

There are 4000 fibers which observed 4000 different objects in a FOV of 5 degrees as mentioned previously. Due to
the FOV of the BATC survey is about 1 degree, most of the fibers lie out of the BATC sky field. In addition, some of the rest
spectra come from point sources such as the foreground stars and planetary nebulae, not from the galaxy itself. Finally,
excluding the spectra with low S/Ns and bad flux calibrations, we totally obtain 59 usable spectra of relatively high qualities
which are located on the M31 surface of our interest.
Figure \ref{fig3} shows the positions of those 59 fibers on the true color map of the Andromeda galaxy which is combined with
three BATC intermediate band images (effective wavelengths of 4550, 6075 and 8082 {\AA} for the blue, green and red channels,
respectively). Most of the spectra are originated from the bulge and bright disk, while a couple of them are related to the dust
lane and spiral arms. For the next analyses of this paper, we smooth the spectra in the whole wavelength range by averaging the
fluxes in every five data points to improve the S/Ns.

\section{Results and Discussions} \label{sec-dis}
\subsection{Parameters Derived by STARLIGHT}
A stellar population synthesis code, STARLIGHT (\citealt{fe05}), is employed to get the physical properties of the M31 spectra.
STARLIGHT considers that a complex stellar population, for example a galaxy, should be a linear combination of a set of simple stellar
populations (SSPs). Optimal match of the observed spectrum with the theoretic model spectra is executed by subtly designed algorithms
to obtain a sequence of parameters (e.g., age and metallicity). During the match, an appropriate extinction, a velocity shift and a
velocity dispersion by Gaussian convolution are added to the model spectrum and so the program also outputs the radial velocity, velocity
dispersion and reddening value. The SSP spectral base we adopt comes from the evolutionary synthesis model of \cite{br03} in the
high revolution version. The SSPs in the base are the same as those of \cite{fe05}, which are generated with the initial mass
function of \cite{ch03}. This base is composed of 45 SSPs encompassing 15 ages between 1 Myr and 13 Gyr and three metallicities
($Z$ = 0.2, 1 and 2.5 $Z_\odot$). Galactic reddening law of \cite{ca89} with $R_{V}$ = 3.1 is served as the dust extinction model.

\begin{figure}
\centering
\includegraphics[width=\textwidth, angle=0]{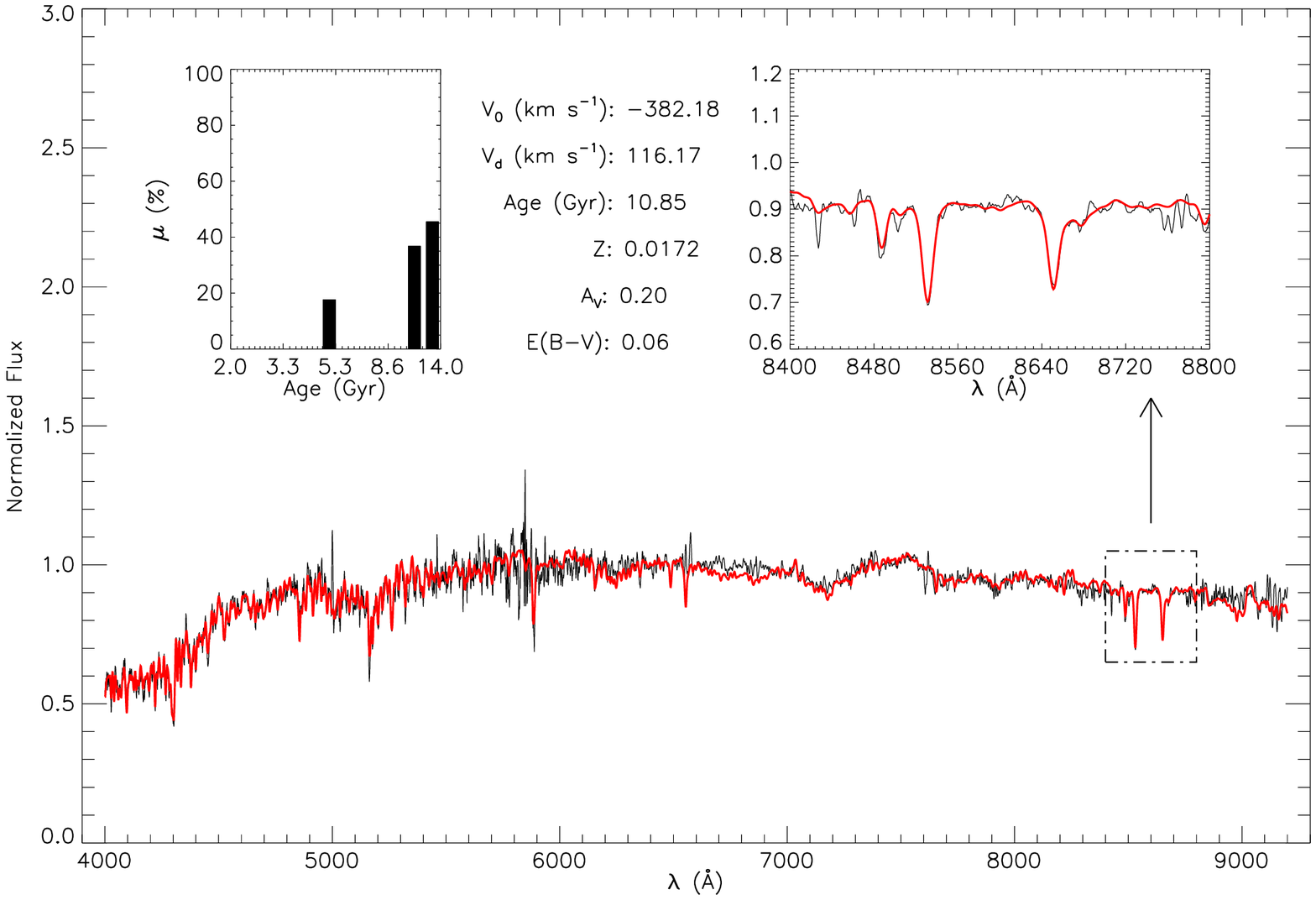}
\begin{minipage}[]{\textwidth}
\caption{Fitting results calculated by STARLIGHT when a spectrum close to the M31 center is considered. The thin black curve
is the observed spectrum and the thick red one is the model spectrum. Both of them are normalized by the flux at 6700 {\AA}.
Detailed observed and model spectra around the Ca {\sc ii} triplet are plotted in the upper right corner. A set of physical
parameter values of the radial velocity ($V_0$), velocity dispersion ($V_d$), age, metallicity ($Z$) and reddening value ($E(B-V)$)
are shown in the middle of this figure. The histogram in black presents the mass weights of different SSPs, revealing the
star formation history (logarithmic x-axis). Note that age and metallicity are the mass-weighted values of all SSP model
spectra.\label{fig4}}
\end{minipage}
\end{figure}

Before being fitted by STARLIGHT, all spectra are corrected by the Galactic extinction of 0.206 in $A_V$ taken from
the reddening map of \cite{sc98}. Besides, they are sampled into integer wavelengths with a step of 1 {\AA}, which is
strongly recommended to do by \cite{fe05}. Figure \ref{fig4} illustrates the fitting result for an individual spectrum
lying in the bulge of M31 ($\alpha = 10^{\circ}.614$, $\delta = 41^{\circ}.249$). The observed spectrum is fitted very
well by the synthetical model spectrum. The plot in the upper right corner of Figure \ref{fig4} displays spectral details
near the Ca {\sc ii} triplet absorption lines (8498, 8542 and 8662 {\AA}) which are partly dependent on the stellar
atmospheric parameters (\citealt{zh91}). Such considerable absorption lines, their profiles and the continuum itself
can ensure the reliable estimation of the radial velocity, velocity dispersion, metallicity and other parameters. In
STARLIGHT, mass weights ($\mu$) of different SSPs can be gained to learn the star formation and chemical abundance
evolution histories. From the histogram in the top-left corner of the Figure \ref{fig4}, we can understand that at that
galactic position, about 82\% of the total mass comes from stellar populations older than 11 Gyr and all the mass was
formed 5 Gyr ago. In addition, the average age of about 11 Gyr and small reddening values of 0.06 mag denote that it
belongs to the typical old bulge stellar population.

\begin{figure}
\centering
\includegraphics[width=\textwidth, angle=0]{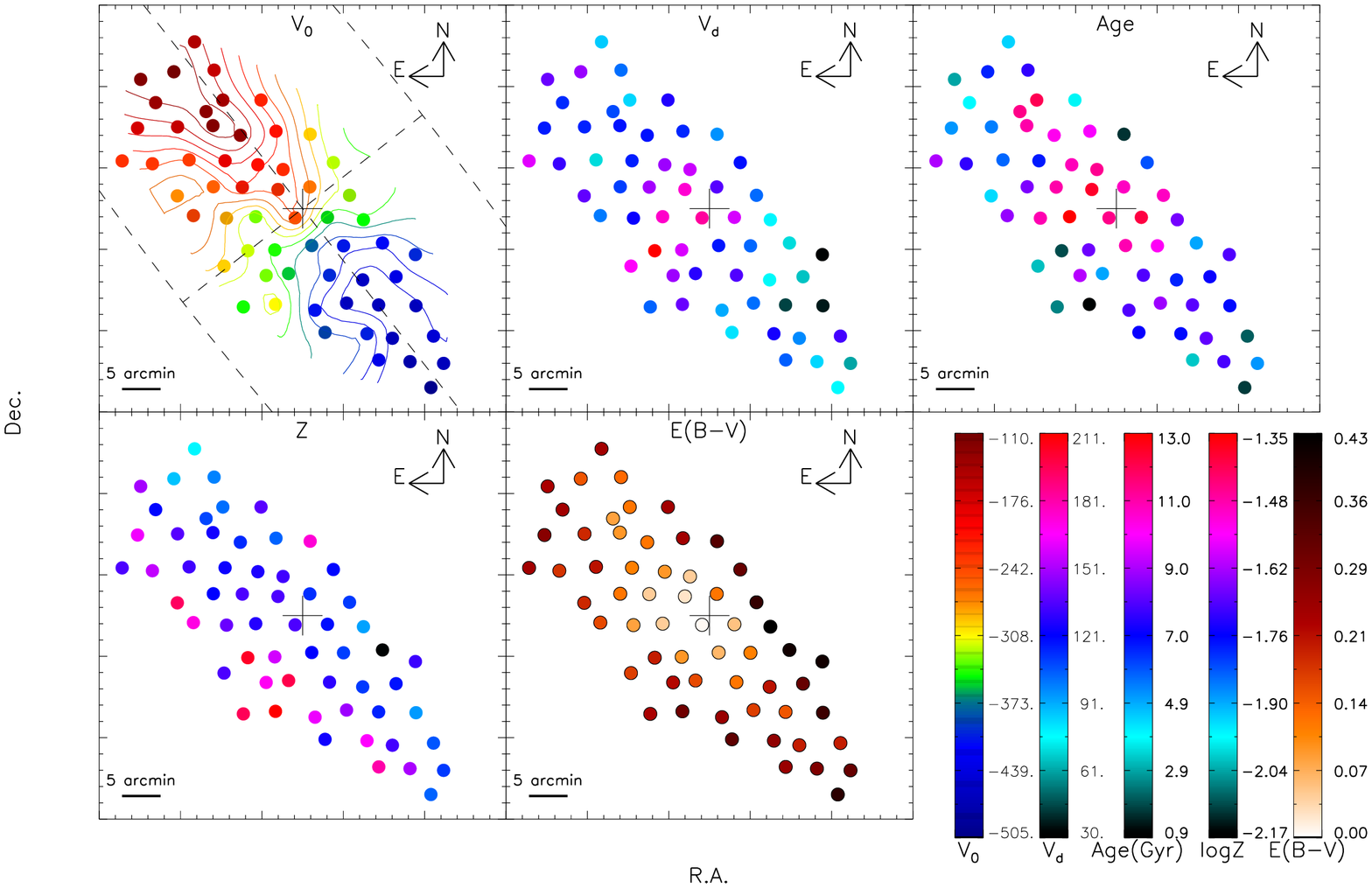}
\begin{minipage}[]{\textwidth}
\caption{Two-dimensional distributions of different parameters derived by STARLIGHT. From left to right and top to bottom,
they are the radial velocity, velocity dispersion, age, metallicity and intrinsic reddening in $E(B-V)$ in their turn. The crossing
symbol in each panel is the optical center ($\alpha$=10.685$^\circ$ and $\delta$=41.269$^\circ$). In the first panel, two
perpendicular dashed lines are the optical major and minor axes. The outer arcs are the clipped elliptical enclosure of
the Andromeda's disk. Here again, we adopt the length of the major axis of about 1.59$^\circ$, the disc inclination angle
of 78$^\circ$ and position angle of the major axis of 38$^\circ$ . Contours are drawn in equally spaced levels within the
age ranges as shown in the color bars. \label{fig5}}
\end{minipage}
\end{figure}

Different parameters for 59 spectra including the radial velocity, velocity dispersion, age, metallicity and extinction
are listed in Table \ref{tab1}. All the radial velocities throughout this paper are converted to the heliocentric frame.
The velocity dispersions are also corrected to true intrinsic stellar dispersions in consideration of the resolution of
the base spectra (3 {\AA} FWHM) and that of the GSJT spectra (R = 1250 as analyzed by \cite{hu10}). Correspondingly, the
two-dimensional maps of these parameters on the galactic surface are displayed in Figure \ref{fig5}.

\begin{table}
\begin{minipage}[]{\textwidth}
\caption[]{Parameters derived by STARLIGHT for all available spectra\label{tab1}}
\end{minipage}
\small
\begin{tabular}{ccccccccccc}
\hline\noalign{\smallskip}
\hline\noalign{\smallskip}
ID & Sp & Fb & R.A.    &Dec.     &V$_0$ &V$_{\rm d}$&Age & $Z$ & $A_V$ & $E(B-V)$ \\
   &    &    & (J2000) & (J2000) &(\kms)  & (\kms)  &(Gyr) &  & mag   & mag      \\
(1) & (2) & (3) & (4) & (5) & (6) & (7) & (8) & (9) & (10) & (11) \\
\hline\noalign{\smallskip}
G2632&15&112&11.1439590&41.5581530& -118.7&  138.7& 2.8&0.0235& 0.71& 0.23\\
G2833&15&107&11.1560830&41.4527510& -174.2&  115.3& 5.1&0.0273& 0.84& 0.27\\
G2932&15&109&11.2027500&41.3821370& -230.6&  160.3& 9.0&0.0200& 0.75& 0.24\\
H2527&15&176&10.9848750&41.6368900& -148.4&   83.7& 4.4&0.0108& 0.76& 0.24\\
H2630&15&187&10.9302500&41.5745280& -163.1&  100.2& 7.5&0.0135& 0.39& 0.13\\
H2631&15&182&11.0471460&41.5730550& -110.2&  147.2& 6.5&0.0119& 0.45& 0.15\\
H2729&15&197&10.7964160&41.5069730& -221.3&  126.2& 3.7&0.0204& 0.76& 0.24\\
H2730&15&188&10.9070840&41.5091100& -151.4&   81.6&12.0&0.0140& 0.38& 0.12\\
H2731&15&192&10.9564340&41.4850010& -125.3&  106.3&11.4&0.0151& 0.24& 0.08\\
H2732&15&119&11.1020000&41.5070000& -133.8&  112.3& 3.9&0.0171& 0.75& 0.24\\
H2828&15&191&10.6569170&41.4298060& -297.6&   94.0& 1.5&0.0302& 1.01& 0.33\\
H2829&15&190&10.7554790&41.4385410& -209.8&  113.0& 9.9&0.0143& 0.74& 0.24\\
H2830&15&196&10.8590000&41.4316670& -139.6&  117.0&10.2&0.0159& 0.37& 0.12\\
H2831&15&178&10.9375620&41.4541400& -118.1&  116.5&11.0&0.0179& 0.29& 0.09\\
H2832&15&193&11.0400210&41.4533040& -163.3&  113.5& 5.4&0.0202& 0.59& 0.19\\
H2927&15&235&10.5920000&41.3670000& -323.3&  117.5& 5.9&0.0170& 0.97& 0.31\\
H2928&15&199&10.7381060&41.3550000& -225.4&  155.9&11.3&0.0195& 0.12& 0.04\\
H2929&15&200&10.8105000&41.3667910& -208.5&  148.1&10.8&0.0182& 0.27& 0.09\\
H2930&15&194&10.9051040&41.3773060& -184.9&  114.6& 6.6&0.0177& 0.33& 0.11\\
H2931&15&198&11.0090000&41.3810000& -235.6&   67.7& 5.7&0.0195& 0.66& 0.21\\
H2932&15&104&11.1153130&41.3744870& -231.7&  130.6& 7.5&0.0248& 0.56& 0.18\\
H3027&15&239&10.5480000&41.2950000& -324.9&  100.7&10.7&0.0154& 1.16& 0.37\\
H3028&15&195&10.6620420&41.3154980& -267.0&  135.5&10.7&0.0158& 0.36& 0.12\\
H3029&15&183&10.7542290&41.3119870& -229.1&  173.9&12.5&0.0200& 0.05& 0.02\\
H3030&15&180&10.8570410&41.3190840& -215.6&  149.9&11.0&0.0200& 0.12& 0.04\\
H3031&15&184&10.9408330&41.3214150& -256.4&  108.8& 8.4&0.0173& 0.36& 0.12\\
H3032&15&185&11.0455840&41.3035850& -277.7&  137.5& 4.3&0.0378& 0.58& 0.19\\
H3127& 3&148&10.5100000&41.2410000& -349.9&   75.8& 8.3&0.0127& 1.33& 0.43\\
H3128& 3&138&10.6135830&41.2485280& -353.9&  158.4&12.2&0.0168& 0.16& 0.05\\
H3129& 4& 44&10.7070000&41.2492490& -245.7&  183.4&11.4&0.0200& 0.00& 0.00\\
H3130& 4& 31&10.8205830&41.2541390& -331.8&  175.0&13.0&0.0185& 0.13& 0.04\\
H3131& 4& 46&10.9049590&41.2526400& -287.8&  121.6&10.9&0.0210& 0.25& 0.08\\
H3132& 4& 26&11.0004160&41.2598320& -241.6&   97.8& 8.7&0.0295& 0.48& 0.16\\
H3226& 3&149&10.3630000&41.1620000& -404.6&   30.5& 7.9&0.0194& 1.26& 0.41\\
H3227& 3&144&10.4569170&41.1893350& -416.1&   68.0& 4.9&0.0067& 1.27& 0.41\\
H3228& 3&133&10.5695620&41.1857360& -412.5&  102.0&10.1&0.0153& 0.33& 0.11\\
H3229& 4& 33&10.6612080&41.1882780& -382.2&  116.2&10.9&0.0172& 0.20& 0.06\\
H3230& 4& 42&10.7680000&41.1810000& -345.6&  159.0& 8.2&0.0257& 0.27& 0.09\\
H3231& 4& 43&10.8459170&41.1807210& -319.2&  210.7& 1.8&0.0418& 0.63& 0.20\\
H3326& 3&137&10.4208340&41.1153340& -440.2&   64.0& 7.0&0.0167& 0.95& 0.31\\
H3327& 3&147&10.5171670&41.1106110& -463.6&   72.8& 6.5&0.0154& 0.65& 0.21\\
H3328& 4& 30&10.6124160&41.1228070& -405.8&  134.3& 7.9&0.0188& 0.38& 0.12\\
H3329& 4& 38&10.7301250&41.1288870& -360.4&  127.1& 4.9&0.0391& 0.51& 0.16\\
H3330& 4& 32&10.7950000&41.1260000& -331.2&  148.3& 9.1&0.0286& 0.69& 0.22\\
H3331& 4& 48&10.9155840&41.1487240& -297.1&  166.9& 3.1&0.0199& 0.53& 0.17\\
H3426& 3&128&10.3660000&41.0510000& -454.2&   35.1& 6.8&0.0129& 1.12& 0.36\\
H3427& 3&134&10.4735410&41.0551680& -469.3&   42.0& 8.0&0.0161& 0.48& 0.15\\
H3428& 4& 34&10.5656660&41.0612790& -455.8&  100.5& 8.8&0.0233& 0.52& 0.17\\
H3429& 4& 47&10.6568750&41.0478060& -413.9&   89.8& 7.9&0.0271& 0.77& 0.25\\
H3430& 4& 28&10.7703750&41.0627520& -310.4&  139.0& 0.9&0.0450& 0.93& 0.30\\
H3431& 4& 40&10.8630000&41.0590000& -347.7&  102.1& 2.4&0.0382& 0.73& 0.23\\
H3526& 3&132&10.3185000&40.9838910& -452.7&  132.2& 2.0&0.0147& 0.61& 0.20\\
H3527& 4&  6&10.4367500&40.9819450& -459.3&   95.2& 8.1&0.0196& 0.61& 0.20\\
H3528& 4& 25&10.5094170&40.9934730& -407.0&  123.8& 6.8&0.0280& 0.81& 0.26\\
H3529& 4& 24&10.6300000&40.9990000& -393.3&   79.5& 7.0&0.0178& 0.99& 0.32\\
H3625& 3&150&10.2920000&40.9240000& -470.6&   58.6& 5.1&0.0154& 0.96& 0.31\\
H3626& 3&140&10.3882910&40.9301380& -481.8&   81.5& 7.8&0.0238& 0.87& 0.28\\
H3627& 4& 23&10.4777500&40.9358600& -435.5&  103.1& 3.2&0.0328& 0.76& 0.25\\
H3726& 3&160&10.3305630&40.8724020& -505.1&   75.1& 1.6&0.0143& 1.19& 0.38\\
\noalign{\smallskip}\hline
\end{tabular}
\tablecomments{\textwidth}{Column (1) is the temporary object ID in the LAMOST catalog. Column (2) and (3) are
the No. of spectrographs and fibers, respectively. Column (4) and (5) give the equatorial coordinates in degrees.
Column (6) is the heliocentric radial velocity. Column (7) is the intrinsic velocity dispersion. Column (8) is
the mass-weighted age. Column (9) is the mass-weighted metallicity. Column (10) is the intrinsic extinction ($A_V$),
and column (11) is the corresponding reddening value in $E(B-V)$ with $R_V$ of 3.1.}
\end{table}

\subsection{Parameter Uncertainties}
For the estimations of the physical parameters, there are several sources of errors. The accuracy of the wavelength
calibration has effect on the determination of the radial velocity. In the wavelength range of the red spectra, there
are a considerable number of night sky emission lines such as OH and O lines. We use about 30 sky lines which are
relatively strong, unblended and uniformly scattered to analyze the accuracy of the wavelength calibration for the
red spectra. This gives the accuracies of about 6 \kms, 4 \kms and 9 \kms for the No. 3, 4 and 15 spectrographs, respectively.
Due to few sky lines in the blue spectra, we crudely estimate the accuracy of the wavelength calibration using
the O{\sc i} line (5577 {\AA}), which gives the accuracy of less than 4 \kms for those three spectrographs. The
accuracy measurements are similar to that of \citet{hu10} who gave the global accuracy of about 8 \kms.

For our flux calibration method, most of the uncertainties should come from the photometric uncertainties in the
BATC 15 intermediate bands. Due to the deep exposures of these bands, the average photometric errors of all the objects
are smaller than 0.05 mag and can reach as low as to 0.02 mag. Such precise photometry makes us believe that the flux
calibrations for all the spectra are accurate enough to get the reliable age, metallicity and
reddening value by fitting the continuum with the synthesis model. From the galactic center to the outer regions, the
spectral S/N gradually degrades, which might cause large uncertainties of parameters, especially the velocity dispersion
in the outer regions. Smoothing the spectra broadens the spectral lines which might cause the velocity dispersion
to be overestimated. As a whole, from the simulations of \cite{fe05}, for the spectra with S/N = 10, the uncertainties
of velocity, dispersion, age in logarithm, $A_V$, metallicity in logarithm are about 8.6 \kms, 12.4 \kms, 0.14 dex, 0.05
mag, and 0.13 dex, respectively. In this study, the S/Ns are larger than 10 for most of the spectra, so we consider
those above uncertainties as the upper uncertainty limits of all our fitting results.

\subsection{Velocity Field and Rotation Curve}
In the first panel of Figure \ref{fig5}, the radial velocity ranging from -505 \kms to -110 \kms is approximately
symmetric relative to both the major and minor axes. The average velocity is about -304 \kms, which is close
to the systemic velocity of about -300 \kms and in very good agreements with the fitted systemic velocity of -304.5 \kms
in the H {\sc i} measurements of \cite{ch09}, although our position distribution of the spectra is a little deviated from
symmetry relative to the rotation center. These velocities presents a rotational velocity field of the stellar
content of M31. They just complements those of the gas content such as H {\sc i} and CO which is scarce near the
core and in the inner disk of the galaxy.

\begin{figure}
\centering
\includegraphics[width=\textwidth, angle=0]{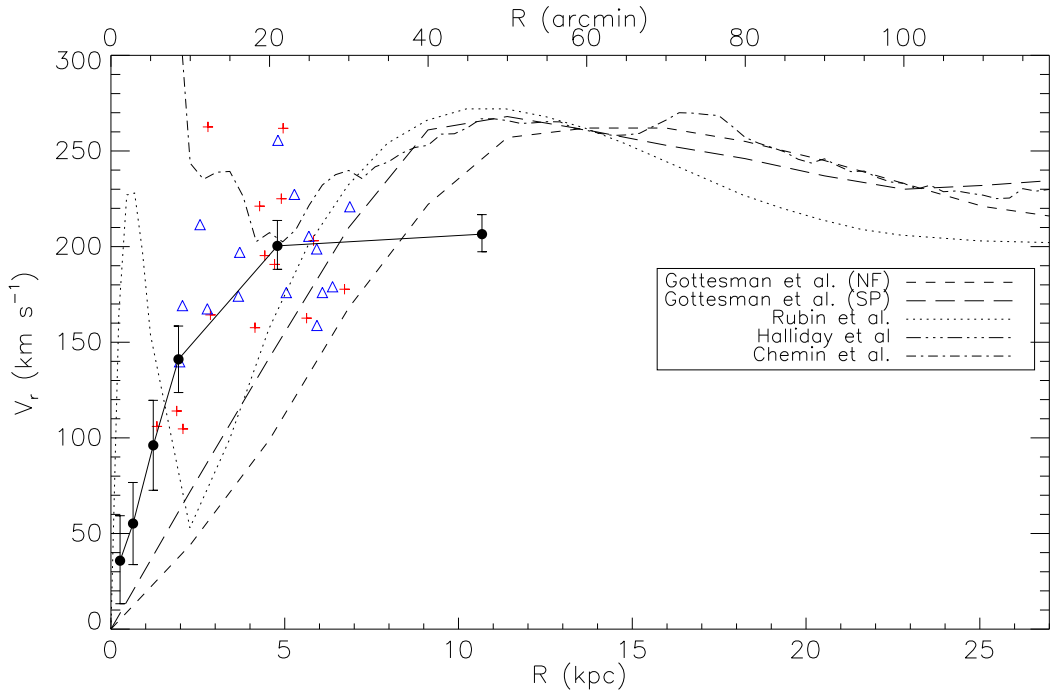}
\begin{minipage}[]{\textwidth}
\caption{Rotation velocity ($V_r$) along the major axis as a function of the radial distance from the center of the Andromeda galaxy.
The crossings are the velocities along the north major axis and the triangles are those along the opposite direction.
The dashed and long-dashed curves are the north following (NF) and south preceding (SP) rotation curves, respectively,
which were determined by the neutral hydrogen observations of \cite{go70}. The dotted curve is the rotation curve
derived by \cite{ru70}. The filled circles with error bars connected with solid lines are the average rotation velocities
in Table 2 of \cite{ha06}. The dash-dotted lines come from the measurements of
\cite{ch09}.
 \label{fig6}}
\end{minipage}
\end{figure}

According to the velocity field and combining other velocity measurements, we can deduce the rotation curve
of the galaxy, which can be used to constrain the galactic potential and the mass distribution and make it
possible in detecting non-circular velocity components caused by either the radial expansion or the asymmetries
due to the disk warp. Figure \ref{fig6} shows the rotation velocity as a function of the radial distance from the
galactic nucleus. The distances and rotation velocities are calculated under the assumption of a pure projected
circular rotation with our selected dynamical parameters as covered previously (distance: 784 kpc, disc inclination:
78$^\circ$, and position angle: 38$^\circ$). In the figure, we only display the velocities of those positions
within 1 kpc away from the major axis as denoted in filled circles in Figure \ref{fig3}. The velocities along the north direction
of the major axis are denoted as crossings and those along the opposite direction are plotted as triangles. We do
not find any discrepancy of the rotation velocity in these two sides, indicating that the stellar content of
this galaxy is rotationally symmetric within about 7 kpc.

We compare our rotation velocities with the published rotation curves, which were derived by H {\sc i} 21-cm
observations of \cite{go70} and \cite{ch09}, emission lines (H$\alpha$ and [N {\sc ii}]) of \cite{ru70} and
planetary nebulae of \cite{ha06}. Our circular velocities are close to other measurements around 5 kpc of
both the gas and stellar contents. However, in the inner range, unlike to curve of \cite{ch09}, the velocity
seems to decrease, which may be due to the rotational difference of those two contents in the galaxy.
As a whole, our rotation velocities are proximate to those of \cite{ha06}.

\subsection{Velocity Dispersion}
In the velocity dispersion distribution of Figure \ref{fig5}, larger dispersions are concentrated on the galactic center,
indicating the bulge is more dynamically thermal. In the east of the galaxy where a spiral arm is located as shown in the
colored map of Figure \ref{fig3}, some of the velocity dispersions become a little larger, which may be caused by the
perturbations of density waves. The global average velocity dispersion is about 114 \kms and the dispersion close to the
nucleus is about 183 \kms. If we take the effective radius of the Andromeda bulge as 282.2 arcsec (\citealt{ba98}), the
average velocity dispersion of the bulge is about 153 \kms, which generates the ratio
of the dispersions between the bulge and nucleus of about 0.84. This ratio is very approximate to that of 0.83 derived
by \cite{wh80}. In addition, the velocity dispersion of the nucleus bulge from \cite{pr78} and \cite{wh80} is about
150 \kms. Global average dispersions from \cite{ha06} and \cite{me06} are about 105 \kms. All these measurements present a
good consistency of the velocity dispersion with our results.

\begin{figure}
\centering
\includegraphics[width=0.8\textwidth, angle=0]{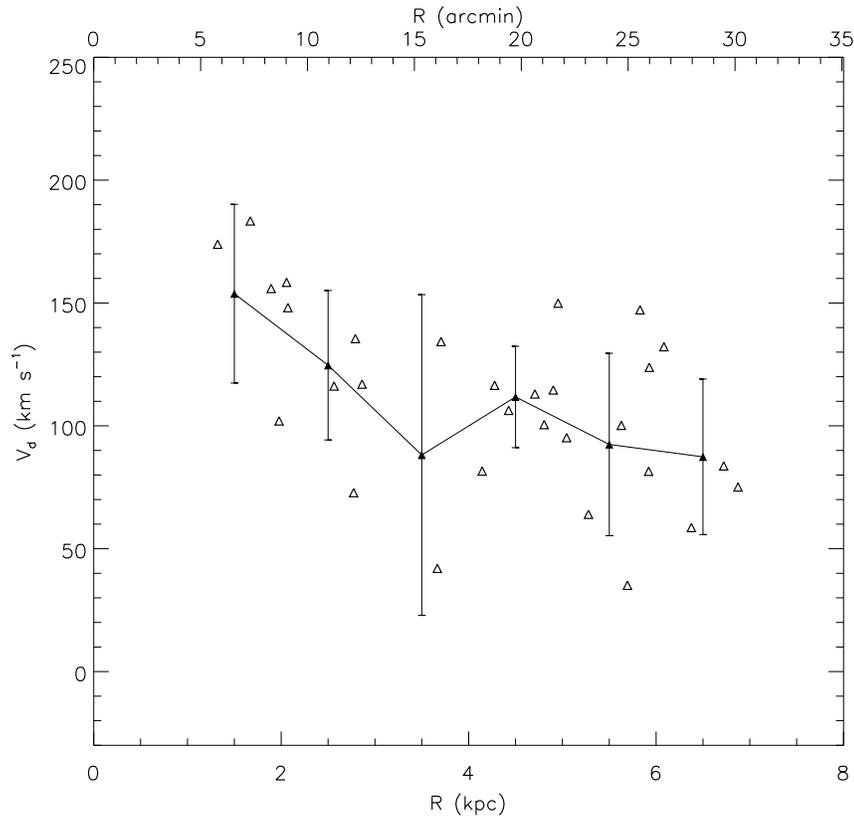}
\begin{minipage}[]{\textwidth}
\caption{Radial distribution of the velocity dispersions along the major axis. The filled triangles with error bars
are their average values within an interval of 1 kpc from 0.0 to 7.0 kpc. The errors are calculated as the standard
deviations.
\label{fig7}}
\end{minipage}
\end{figure}

Figure \ref{fig7} presents the radial distribution of the velocity dispersion along the major axis. The radial
distances of all positions adopted here and in the rest of our paper are corrected with the characteristic parameters
of the galaxy as previously mentioned. We can see that the dispersion becomes smaller when it goes further away
from the center. The variation tendency is consistent with with the results of \cite{ha06} and \cite{me06} in
their study of PNe. As the distance extends for 4 kpc further, the dispersion seems to become stable around 100 \kms.

\subsection{Age, Metallicity and Extinction}
From the age distribution in Figure \ref{fig5}, we can see that the bulge is older and younger stellar populations are likely
to be located on the disk and near the spiral arms, which can be confirmed by the narrow band observations of emission lines. The
average age is about 7.3 Gyr and the age of the bulge is about 11.5 Gyr showing the oldest component of the galaxy. The stellar
populations close to the spiral arms can reach as young as about 1 Gyr. Such young components are likely to associate with the
H {\sc ii} regions where a number of young massive stars are being born.

\begin{figure}
\centering
\includegraphics[width=0.8\textwidth, angle=0]{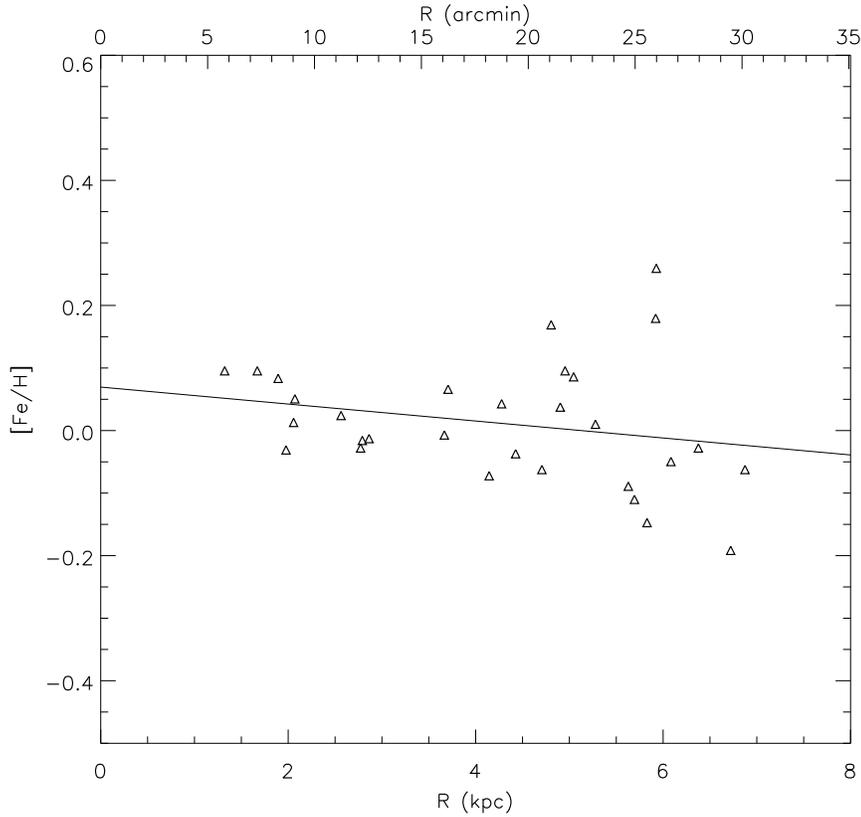}
\begin{minipage}[]{\textwidth}
\caption{Radial distribution of the metallicity in [Fe/H] along the major axis. The solid line shows the
linearly fitted metallicity gradient. \label{fig8}}
\end{minipage}
\end{figure}

Abundance map in Figure \ref{fig5} shows that the spiral arms in the east are richer than other parts of the galaxy. The
global average metallicity is about 0.02, the same as the solar abundance. For all spectra possible from the spiral arms
and near the H {\sc ii} regions, their mean metallicity is about 0.032, which is richer than the solar metallicity.
We convert the metallicity in $Z$ into [Fe/H] with the solar chemical composition of \cite{gr98}. Although the scatter
in this sample is substantial, the radial [Fe/H] distribution along the major axis as presented in Figure \ref{fig8}
gives a small metallicity gradient of -0.014 dex kpc$^{-1}$.

\cite{bl82} determined both the nitrogen and oxygen abundances for 11 H {\sc ii} regions by the empirical method. They
derived the oxygen abundance gradient of about -0.44 dex which was scaled to the photometric radius of R25 where the
surface brightness of the galaxy becomes 25 mag arcsec$^{-2}$. Considering R25 of about 1.59$^\circ$ and the distance
of about 784 kpc, the gradient of -0.44 dex in \cite{bl82} is equal to -0.02 dex kpc$^{-1}$, which is somewhat larger
than our result. The mean oxygen abundance of those 11 H {\sc ii} regions is about 8.81 in log(O/H)+12. This gives
the metallicity $Z$ of about 0.016 which is a little lower than the average metallicity of 0.02 in our work.
Globular clusters were gathered up to give the metallicity distribution by \cite{ba00} and \cite{fa08}, who showed
a bimodal profile with peaks of [Fe/H] = -1.4 ($Z = 0.0007$) and [Fe/H]= -0.6 ($Z = 0.004$) for the poor and rich
groups, respectively. These globular clusters as the fossils during the evolution of the galaxy show much lower
abundances than those of the galactic bulge and disk stellar populations as we measure.

The reddening distribution in Figure \ref{fig5} shows that the bulge is clear of dust, while large extinctions tend to lie on the dust
lane and the spiral arms, which take on a dusty ring. The average reddening is about 0.20 in $E(B-V)$. \cite{ku79} measured the
reddening values of 22 H {\sc ii} regions and presented the mean $A_V$ of about 1.31 (0.42 in $E(B-V)$), which is much larger
than the average of our measurements but closer to those of the spiral arms and H {\sc ii} regions. Both \cite{ba00} and
\cite{fa08} determined the reddening values of globular clusters in M31 using the correlations between optical and infrared
colors and metallicity as well as by defining various ``reddening free" parameters. They gave the mean reddening values
of about 0.22 and 0.28, respectively, close to our results. They also reported that the reddening is larger in the northwest
of the galaxy than the other side, which can also be indicated in the reddening map as shown in the last panel of Figure \ref{fig5}.

\subsection{Star Formation History}
\begin{figure}
\centering
\includegraphics[width=0.8\textwidth, angle=0]{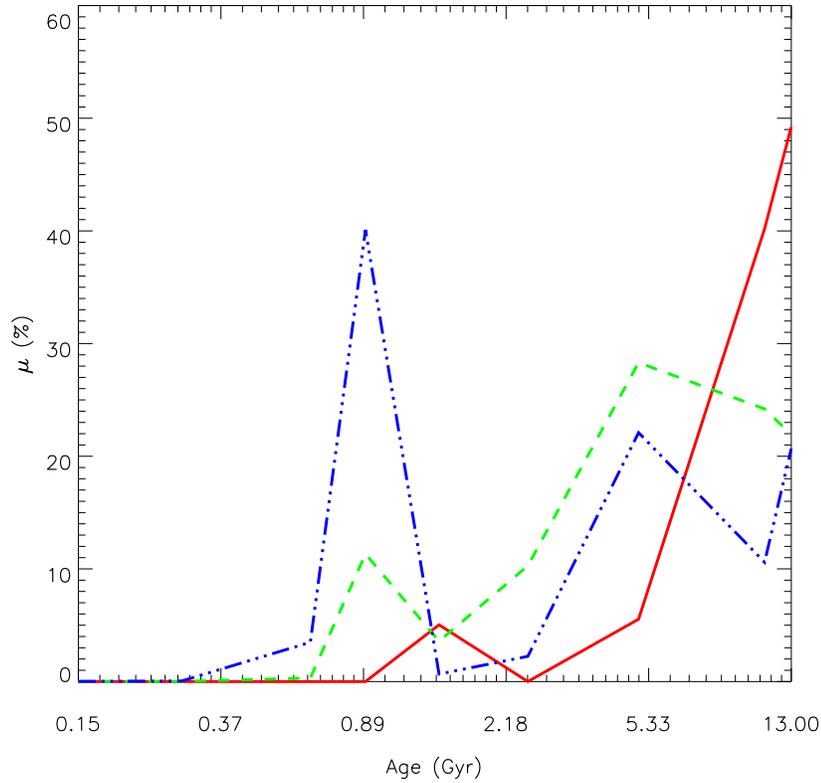}
\begin{minipage}[]{\textwidth}
\caption{Average mass fractions of the SSPs with the same ages for the three different components of the galaxy.
The red solid lines are the mass fractions for the bulge, the green dashed ones are those for the disk and the blue dash-dotted
ones are those for the spiral arms. The abscissa is logarithmic.
\label{fig9}}
\end{minipage}
\end{figure}

We divide all the spectra into three different components, i.e., the bulge, disk and spiral arms. Five spectra in total are
located in the bulge and ten spectra lies in the spiral arms where much younger and metal-richer stellar populations
are present. The rest of the spectra are considered to reside in the disk. For a specified spectrum, the
mass fractions of SSPs with different ages are provided by STARLIGHT during fitting the spectrum with a linear combination
of the SSP models. These mass fractions of different ages give the stellar evolution history. We plot the average mass
fraction of the SSPs with the same age for each components in Figure \ref{fig9}. Most of the stellar mass in the bulge
was formed as early as about 10 Gyr and no more star formation has occurred recently in this region. The disk seems to
be more continuous in forming new generations of stars in its history and most of the disc stellar mass was formed about
5 Gyr ago. The spiral arms not only contain similar stellar populations of intermediate and old ages as the disk, but also
have considerable young stellar populations formed about 1 Gyr ago. Since the spiral arms are the places where material
density is perturbed by the density waves of the spiral structures, there are a great number of young massive stars are being born.

\section{Conclusion} \label{sec-conc}
The GSJT is a fiber-feeding spectroscopic telescope with both a large field of view and a large aperture, which can
obtain 4000 fiber spectra simultaneously. During the 2009 commissioning phase, two M31 fields were observed: one is
centered near the galaxy core and the other is located in the halo. Since the data processing pipelines for the GSJT
observed data are not in their perfect status and no suitable standard stars are found for flux calibration, we use
the photometric data of 15 intermediate bands in the BATC survey which also observed M31 to flux-calibrate all
the spectra in the field focused on the M31 center. As a result, there are 59 usable spectra in total. We use these
spectra to study the kinematic properties and stellar populations of this galaxy. By STARLIGHT, we obtain the radial
velocities, velocity dispersions, ages, metallicities and reddening values of all those 59 spectra. The distributions
of these parameters are presented and comparisons with other measurements are also performed. The main conclusions
are summarized as below:

(1) The radial velocity ranges from -505 to -110 \kms, extending to the distance of about 7 kpc along the major axis.
The average velocity is about -304 \kms, very close to the systemic velocity of 300 \kms. Rotation velocities of the
spectra are calculated and compared with other rotation curves derived by the observations of the atomic hydrogen,
ionized hydrogen regions and planetary nebulae.

(2) The average velocity dispersion is about 114 \kms. The dispersions close to the nucleus and of the bulge within
R = 282.2{\arcsec} are 183 \kms and 153 \kms, respectively, which yield the ratio of about 0.84 between the bulge and
nucleus. This ratio approximates that of 0.83 derived by \cite{wh80}. The radial dispersion distribution shows that
the dispersion becomes smaller as it is away from the center and the bulge is more dynamically thermal than the disk.

(3) The average age is about 7.3 Gyr and the bulge was formed about 11.5 Gyr ago. Some places close to the spiral
arms regions can be as young as about 1 Gyr. The spiral arms and the areas close to H {\sc ii} regions are both
youngest and richest in abundance.

(4) The average abundance $Z$ of the stellar content is about 0.02, the same as the solar metallicity. A small radial
gradient of about -0.014 dex kpc$^{-1}$ is gained. This metallicity gradient of the stellar content are somewhat lower
than those of the gas content in the paper of \cite{bl82} who determined the abundances of 11 H {\sc ii} regions.

(5) The reddening map shows that the nucleus and bulge is clear of dust and a distinct dust ring surrounds the galactic center.
The mean reddening is about 0.2 in $E(B-V)$ which approximates the average reddening value of the M31 globular clusters.

(6) The star formation history presents that the bulge, disk and spiral arms were formed in different stages of the evolution history.
Most the stellar mass of those three components were formed 10 Gyr, 5 Gyr and 1 Gyr ago.

\begin{acknowledgements}
The Guoshoujing Telescope (GSJT) is a National Major Scientific Project built by the Chinese Academy of Sciences. Funding for the
project has been provided by the National Development and Reform Commission. The GSJT is operated and managed by the National
Astronomical Observatories, Chinese Academy of Sciences. The STARLIGHT project is supported by the Brazilian agencies CNPq, CAPES
and FAPESP and by the rance-Brazil CAPES/Cofecub program. This work was supported by the Chinese National Natural Science
Foundation grands No. 10873016, 10633020, 10603006, 10803007, 10903011, 11003021, and 11073032, and by National Basic Research Program of
China (973 Program), No. 2007CB815403.
\end{acknowledgements}

\label{lastpage}

\end{document}